# ENHANCING ALGEBRAIC QUERY RELAXATION WITH SEMANTIC SIMILARITY

Lena Wiese

*University of Hildesheim - 31141 Hildesheim, Germany*

**ABSTRACT**

Cooperative database systems support a database user by searching for answers that are closely related to his query and hence are informative answers. Common operators to relax the user query are Dropping Condition, Anti-Instantiation and Goal Replacement. In this article, we provide an algebraic version of these operators. Moreover we propose some heuristics to assign a degree of similarity to each tuple of an answer table; this degree can help the user to determine whether this answer is relevant for him or not.

**KEYWORDS**

Query relaxation, query generalization, similarity degree, relational algebra

## 1. INTRODUCTION

In traditional database systems users either retrieve exact answers to their queries or no answers at all; that is, the result table either contains all exactly matching tuples or no tuples at all. In case no answer is found, the query is said to be a "failing query" (see for example, Pivert 2010). In contrast, so-called *cooperative database systems* provide users with *informative answers*: when no exact answer can be found, the database system tries its best to return a related answer that might satisfy the user's intent at least to some degree. One of the earliest works for the handling of failing queries is the CoBase system of Chu et al (1996) that relies on generalizations of constants in type abstraction hierarchies. Other related systems like Flex (Motro 1990), Carmin (Godfrey et al 1994) and Ishmael (Godfrey 1997) introduce generalization operators dedicated for logical queries. Hurtado et al (2008) relax RDF queries based on an ontology. In a multi-agent negotiation framework, Sakama and Inoue (2007) devise procedures for generating "neighborhood proposals" and "conditional answers" in a negotiation process. Similarly, Sá and Alcantara (2011) restrict query relaxation by providing rankings on what they call explanations for the failure of a query (but without proposing a concrete preference relation) and marking some literals as non-replaceable. Halder and Cortesi (2011) employ abstraction of domains and define optimality of answers with respect to some user-defined relevancy constraints. The approach by Pivert et al (2010) using fuzzy sets analyzes cooperative query answering based on semantic proximity. With what they call incremental relaxation they apply generalization operators to single conjuncts in a conjunctive query; hence generalized queries form a lattice structure: queries with the same number of applications generalization operators are in the same level of the lattice. In the work of Inoue and Wiese (2011) and the corresponding CoopQA system (http://www.coopqa.de), cooperative database behavior is obtained by logically generalizing queries and returning answers to these generalized queries as informative answers. Its main result is that three generalization operators for conjunctive queries can be applied iteratively and it is sufficient to apply these three operators in a certain order (hence avoiding the need to compute their reverse iteration).

Most of these systems analyze the behavior of their operators for conjunctive queries: a conjunctive query consists of a conjunction of logical literals. Three common operators for the logical generalization of conjunctive queries are Dropping Condition, Anti-Instantiation and Goal Replacement:

- Dropping Condition (DC) removes one conjunct from a query and hence looks for answers that satisfy less conditions than the original query.

- Anti-Instantiation (AI) replaces a constant (or a variable occurring at least twice) in the original query with a new variable and hence looks for answers that have less restrictions in terms of concrete values for or equality constraints on variables at a certain position.
- Goal Replacement (GR) applies a deduction rule to the query; the rule can be a database constraint or more generally be an element of a knowledge base Σ on which the query is evaluated. A rule consists of a body part left of an implication arrow (→) and a head part right of the implication arrow. The body of a consists of a conjunction of literals whereas the head is often restricted to consist only of one single literal; additionally some form of "range-restriction" for variables is required. The GR operator then finds a substitution $\theta$ that maps the rule's body to some conjuncts in the query and replaces these conjuncts by the head (with $\theta$ applied).

In many logic-based cooperative database systems (for example in Inoue and Wiese (2011)), after each generalization step, all informative answers are returned in no particular order although some answers might have higher relevancy than others. In other papers (for example, Halder and Cortesi (2011) and Sá and Alcantara (2011)), relevancy or preference orderings are supposed to be given as input by the users; however the authors do not provide means or methods to obtain such orderings. In this article, we hence propose some improvements and make the following contributions to cooperative information systems:

1. We provide a restatement of the logical generalization operators in terms of relational algebra and hence provide a first step to use the generalization operators in a relational database system.
2. We define similarity-based weighting strategies for query generalization that can be used to define rankings or preferences of informative answers and hence avoid the need for user-defined orderings. The weights hence correspond to some form of semantic relevancy of informative answers.

The rest of the paper is organized as follows: Section 2 covers the background on cooperative query answering and generalization and provides the translation of the logical operators into algebraic expressions. Section 3 presents several heuristics that can be used to determine the relevancy of a tuple in an answer table; in particular, similarity between constants is used to define such relevancy degrees. Section 4 concludes the article.

## 2. QUERY GENERALIZATION WITH RELATIONAL ALGEBRA

As is common in the related work, we will focus on conjunctive queries which are defined in a logical setting based on literals that are combined by logical conjunction and possibly prefixed by existential quantifiers binding some of the variables appearing in the literals. A literal contains an atomic formula consisting of the relation symbol of a database relation with its parameters of variables and constants and possibly a negation symbol.

**Definition 1 (Conjunctive query).** *A conjunctive query is a conjunctive formula $L_1 \wedge \ldots \wedge L_n$ where each $L_i$ is a literal. It can have a prefix of existential quantifiers ∃ that bind some of the variables in the $L_i$. We often abbreviate a query as Q(x), where Q stands for the conjunction of literals (and possible the existential prefix) and x is a tuple of variables appearing freely in Q.*

We illustrate conjunctive queries with the following running example. Assume a database for a medical record with two tables: one table named *Ill* and two attributes *Name* and *Disease* that links patients' names to their diseases; and another table named *Treat* that records prescriptions of patients with the two attributes *Name* and *Prescription*. As an example, *Q(x) = Ill (x,Flu)* ∧ *Ill (x,Cough)* asks for all the names *x* of patients who suffer from both flu and cough as recorded in the medical record. Note that in general the generalization operators DC, AI and GR are fully applicable also to queries containing negative literals but query answering on unrestricted negative literals may lead to undesirable behavior (like infinite answers) depending on the theoretical background model involved. That is why in the context of this article we restrict ourselves to positive conjunctive queries that do not allow negative literals.

### 2.1 Expressing conjunctive queries in relational algebra

Positive conjunctive queries can be translated into Selection-Projection-Join queries in the classical relational data model. We provide some examples in Table 1.

Table 1. Correspondence between logic and algebraic expressions for positive conjunctive queries.

| Conjunctive query | Algebraic query |
|---|---|
| $Ill(x,Flu)$ | $\pi_{Name}(\sigma_{Disease=Flu}(Ill))$ |
| $Ill(x,Cough)$ | $\pi_{Name}(\sigma_{Disease=Cough}(Ill))$ |
| $Ill(x,Flu) \wedge Ill(x,Cough)$ | $\pi_{Name}(\sigma_{Disease=Flu}(Ill)) \bowtie \pi_{Name}(\sigma_{Disease=Cough}(Ill))$ |
| $Ill(y,Flu) \wedge Ill(x,Cough)$ | $\pi_{Name}(\sigma_{Disease=Flu}(Ill)) \times \pi_{Name}(\sigma_{Disease=Cough}(Ill))$ |
| $Ill(x,y) \wedge Ill(x,Cough)$ | $\pi_{Name,Disease}(Ill \bowtie (\sigma_{D2=Cough}(\rho_{D2 \leftarrow Disease} Ill))$ |
| $Ill(x,Flu) \wedge Ill(x,y)$ | $\pi_{Name,Disease}((\sigma_{D1=Flu}(\rho_{D1 \leftarrow Disease} Ill))$ |
| $Treat(x,Inhaler) \wedge Ill(x,Cough)$ | $\pi_{Name}((\sigma_{Prescription=Inhaler} Treat) \bowtie (\sigma_{Disease=Cough} Ill))$ |

In general, for positive conjunctive queries, their algebraic representation corresponds to an equi-join (with equality conditions according to variables shared between literals) on the relation symbols contained in each literal; if attribute names are not disjoint, they must be renamed before the join. After the join, selection has to take place according to the constants appearing in the literals. And finally a projection on the attributes corresponding to free variables in the literals is executed. So, a general algebraic expression for conjunctive queries $L_1 \wedge ... \wedge L_n$ is the following: Let $R_i$ be the relation symbol occurring in literal $L_i$, $A$ the set of attributes corresponding to free variables in the query $Q$ (if a free variable occurs for more than one attribute, one of the attributes is chosen arbitrarily), $C$ the set of equalities $A_j=a_j$ where $a_j$ is a constant occurring in the query for attribute $A_j$, $E$ the set of equalities between attributes $A_k=A_l$ where for attributes $A_k$ and $A_l$ the same variable occurs in the query and lastly a set $R$ of renaming conditions $A_{ij} \leftarrow A'_{ij}$ in case the joined relations contain identical attribute names, then the algebraic expression for the query is:

$$\pi_A[\sigma_C(\rho_R(R_1) \bowtie_E ... \bowtie_E \rho_R(R_n))]$$

For the example query $Q(x) = Ill(x,Flu) \wedge Ill(x,Cough)$, the resulting algebraic notation would be as follows where also renaming $\rho$ is used to differentiate the duplicate use of $Ill$:

$$\pi_{N1}\sigma_{D1=Flu,D2=Cough}(\rho_{N1 \leftarrow Name, D1 \leftarrow Disease}(Ill)) \bowtie_{N1=N2} \rho_{N2 \leftarrow Name, D2 \leftarrow Disease}(Ill))$$

Now, assume the following example database with the two tables $Ill$ and $Treat$:

$Ill$:

| Name | Disease |
|---|---|
| Mary | Cough |
| Mary | BrokenLeg |
| Mary | Sinusitis |
| Pete | Flu |

$Treat$:

| Name | Prescription |
|---|---|
| Mary | Inhaler |

The example query $Q(x)$ is failing in the example database: there is no tuple in the resulting table. Note that it might be more efficient to execute the selection $\sigma$ before executing the join as shown in Table 1.

## 2.2 Transforming generalization operators into algebra

We now describe how the three generalization operators DC, AI, and GR affect the algebraic expression.

- Dropping conditions corresponds to dropping one relation $R_i$ (including the renaming $\rho_R$) from the join expression and appropriately adding or removing conditions from $A$, $C$ and $E$.
- Anti-Instantiation corresponds to adding an attribute (the one for which the new variable was introduced in the logical variant of the query) to the set $A$ of attributes on which projection takes place and removing any condition for this attribute from the selection conditions $C$ (if a constant was anti-instantiated) or from the equality conditions $E$ (if a variable was anti-instantiated).
- Goal replacement corresponds to replacing (according to the rule chosen) several of the relations $R_i$ from the join expression with a single new relation (including the appropriate renaming $\rho$) and appropriately adding or removing conditions from $A$, $C$ and $E$.

We illustrate the three operators with our running example. Dropping conditions on the example query results in two queries: $\pi_{N1}\sigma_{D1=Flu}(\rho_{N1 \leftarrow Name, D1 \leftarrow Disease}(Ill))$ with the single answer tuple *Pete* for the name attribute $N1$, as well as $\pi_{N2}\sigma_{D2=Cough}(\rho_{N2 \leftarrow Name, D2 \leftarrow Disease}(Ill))$ with the single answer tuple *Mary* for the name attribute $N2$. Anti-Instantiation results in three queries:

$\pi_{N1,N2}\sigma_{D1=Flu,D2=Cough}(\rho_{N1 \leftarrow Name, D1 \leftarrow Disease}(Ill) \bowtie \rho_{N2 \leftarrow Name, D2 \leftarrow Disease}(Ill))$ with the answer table

| N1 | N2 |
|---|---|

$\pi_{N1,D1}\sigma_{D2=Cough}(\rho_{N1\leftarrow Name,D1\leftarrow Disease}(Ill) \bowtie_{N1=N2} \rho_{N2\leftarrow Name,D2\leftarrow Disease}(Ill))$ with the answer table

| N1 | D1 |
|---|---|
| Mary | Cough |
| Mary | BrokenLeg |
| Mary | Sinusitis |

$\pi_{N1,D2}\sigma_{D1=Flu}(\rho_{N1\leftarrow Name,D1\leftarrow Disease}(Ill) \bowtie_{N1=N2} \rho_{N2\leftarrow Name,D2\leftarrow Disease}(Ill))$ with the answer table

| N1 | D2 |
|---|---|
| Pete | Flu |

Goal replacement with the rule $Ill(x,Flu) \rightarrow Treat(x,Inhaler)$ results in one query:

$\pi_{N1}\sigma_{P1=Inhaler,D2=Cough}(\rho_{N1\leftarrow Name,P1\leftarrow Prescription}(Treat) \bowtie_{N1=N2} \rho_{N2\leftarrow Name,D2\leftarrow Disease}(Ill))$ with the answer table

| N1 |
|---|
| Mary |

As shown by the above examples, the three operators DC, AI and GR help with finding answers that provide related information to a user who's original query might be a failing one. However, it is often unclear which position or conjunct to choose for generalization; or with which generalization to start in case several operators are applied iteratively. This issue will be handled in the next section by a similarity-based weighting strategy.

## 3. SIMILARITY-BASED WEIGHTING OF ANSWERS

Given a failing query, the question arises which application of a generalization operator retrieves the best informative answers for the user. For AI we have to choose an attribute occurring in the selection conditions $C$ or the equality conditions $E$; for DC, any relation $R_i$ can be removed from the join expression; and for GR potentially several rules can be applied. For each such choice, we will determine a weight considering similarity of constants or loss of information such that only answers with an optimal weight can be returned to the user. We analyze each of the operators in the following subsections.

### 3.1 Anti-instantiation

For AI, any attribute occurring in the selection conditions $C$ or the equality conditions $E$ can be chosen. The selection conditions $C$ consist of expressions of the form $A_j=a_j$ where $a_j$ is a constant and $A_j$ an attribute. We let a notion of similarity between constants induce the weight for each answer tuple.

For similarity of a pair of constants $a_j$ and $b$ several cases arise:
1. For the case that $a_j$ is a proper noun (like *Flu* or *Cough*) there exist ontologies and corpora like *WordNet* that provide a taxonomy of proper nouns. Similarity values (for example, a degree of similarity as a value from the unit interval $[0...1]$) for each pair of proper nouns can be obtained with methods proposed by Leacock and Chodorow (1998), Wu and Palmer (1994), or Resnik (1995).
2. For non-proper nouns (like names of persons *Mary* or *Pete*), we propose to use a notion of acquaintance or geographical closeness of persons, locations et cetera. For example, acquaintance of people can be determined by their connection in a social network or a family tree (using similar methods as for proper nouns to obtain similarity values in a taxonomy). For our running medical example, this can be justified by the fact that if Mary and Pete are friends or close family members, the probability that they suffer from the same infectious disease might be higher than for unrelated persons.
3. For numerical values, a scaling function can be used (as proposed by Belohlavek et al (2011)) to obtain a degree of change over the whole range of numbers that can occur for the constant $a_j$. For example, if we have a price attribute ranging from 0 to 1000, and we are interested in the similarity of the two values $a_j=100$ and $b=110$, then we can divide the difference of the two values by the length of the whole range and subtract this from 1; that is, the similarity of 100 and 110 would be $1 - (|110 - 100|/1000)=0.99$.

Hence, we assume that a similarity degree *sim* from the unit interval is given for each pair of constants $a_j$ and $b$ that can occur for attribute $A_j$: $sim(a_j,b)$ is a value in *[0...1]*. If now during Anti-Instantiation the expression $A_j=a_j$ is deleted from *C*, then, instead of $a_j$, other constants $b$ for attribute $A_j$ may occur in tuples of the answer table. Similar to the method used by Belohlavek et al (2011), we can assign the similarity degree $sim(a_j,b)$ to each tuple of the answer table where $b$ occurs for attribute $A_j$. In our medical example, when choosing to anti-instantiate on the constant *Flu*, we would have to obtain a similarity degree between *Flu* and *Cough*, *Flu* and *BrokenLeg*, and *Flu* and *Sinusitis*. Assume $sim(Flu,Cough)=0.8$, $sim(Flu,BrokenLeg)=0.4$, and $sim(Flu,Sinusitis)=0.9$. Then we could withhold answers with a low similarity degree. For example, the disease *BrokenLeg* could be removed from the answer table:

$\pi_{N1,D1}\sigma_{D2=Cough}(\rho_{N1\leftarrow Name,D1\leftarrow Disease}(Ill) \bowtie_{N1=N2} \rho_{N2\leftarrow Name,D2\leftarrow Disease}(Ill))$

| N1 | D1 | |
|---|---|---|
| Mary | Cough | 0.8 |
| ~~Mary~~ | ~~BrokenLeg~~ | ~~0.4~~ |
| Mary | Sinusitis | 0.9 |

Moreover, it might happen, that we have to decide, whether it is better to anti-instantiate on *Flu* (as shown above) or on *Cough* as shown below.

$\pi_{N1,D2}\sigma_{D1=Flu}(\rho_{N1\leftarrow Name,D1\leftarrow Disease}(Ill) \bowtie_{N1=N2} \rho_{N2\leftarrow Name,D2\leftarrow Disease}(Ill))$

| N1 | D2 | |
|---|---|---|
| Pete | Flu | 0.8 |

This decision can also be guided by the similarity degrees. An aggregation function can be used to condense the values in the answer table into one and then compare the values for the two answer tables. For example, if we are interested in a maximum degree of similarity, the first answer table is preferred because 0.9 is the maximum value contained in any answer table. If however, we are interested in a good answer on average, then we see that the first answer table (assuming that *BrokenLeg* is not removed) has an average similarity degree of 0.7 while the second answer table has an average of 0.8. A row of other such heuristics could be used to compare quality of two answer tables.

The second case for AI is that an equality condition $A_k=A_l$ from *E* is removed. In this case, similarity degree of a tuple in an answer table can be obtained by determining similarity between the constants for $A_k$ and $A_l$ in the tuple. In the medical example, we would have to obtain a similarity degree for *Mary* and *Pete*, because the equality condition *N1=N2* was removed from *E*. Assuming that they are children in the same family, they could have a similarity value of 0.9 hence resulting in an answer tuple with this degree assigned.

$\pi_{N1,N2}\sigma_{D1=Flu,D2=Cough}(\rho_{N1\leftarrow Name,D1\leftarrow Disease}(Ill) \bowtie \rho_{N2\leftarrow Name,D2\leftarrow Disease}(Ill))$

| N1 | N2 | |
|---|---|---|
| Pete | Mary | 0.9 |

## 3.2 Dropping condition

For Dropping Condition several options for the decision which relation to remove from the join expression arise. First of all, this decision can be independent from any semantic content of the query – assuming that such semantic content is expressed in selection conditions *C* or equality conditions *E*; instead, the decision can purely concentrate on syntactic elements of the whole query or the dropped condition alone. For example, we can conjecture that predicates with less attributes (predicates with a lower arity) convey less information and hence it is preferable to drop them. By dividing the arity of the dropped relation by the sum over all arities of all predicates occurring in the query and then subtracting this fraction from 1, we can obtain a purely syntactic similarity degree. More precisely, if $R_i$ is the relation to be removed, then we can subtract from 1 the ratio of the arity $ar(R_i)$ with respect to the sum of all arities of predicates taking part in the join as $1 - (ar(R_i)/ \Sigma_{j=1...n} (ar(R_j)))$. Each tuple of the answer table of the query where $R_i$ is removed then has this value as its similarity degree. In the running example, as only the *Ill* relation is involved twice in the join, all answer tuples would have degree 0.5.

When deciding which condition to drop, we could also penalize the fact that while dropping a condition, attributes are removed from the projection expression $\pi_A$. In particular, this means that the answer table of the generalized query contains less columns. In this case, dropping a condition where less attributes are removed from the projection is preferred. Additionally or alternatively, we could also analyze the number of variables and constants in the dropped relation: during DC it might be necessary to remove selection or equality conditions from *C* and *E*. Computing the ratio of dropped conditions, we can use this as the penalty for the answer tuples. Subtracting the ratio from 1 results in a similarity degree as in the cases above.

To obtain a semantic version of DC – which implies that potentially all tuples in an answer table have different similarity degrees – we could in analogy to the AI case use aggregation of *sim* values between constants as follows. Assume that during DC a selection condition $A_j=a_j$ is dropped. For each answer tuple let *b* be a constant occurring in this answer tuple. We obtain the similarity between $a_j$ and *b*, $sim(a_j,b)$. We then aggregate these similarity values for all constants *b* occurring in the tuple; again, we could choose the maximum, the average or any other aggregation function. This aggregated value is then the similarity degree for the tuple under consideration. With this mechanism we ensure that those tuples in an answer table get assigned a higher similarity degree the constants of which are more related to the constants in the dropped condition; that is the tuple retains the topic of the dropped condition better. If for example, the query

$$\pi_{D1}\sigma_{D2=Cough}(\rho_{N1\leftarrow Name, D1\leftarrow Disease}(Ill) \bowtie_{N1=N2} \rho_{N2\leftarrow Name, D2\leftarrow Disease}(Ill))$$

(note that the selection condition for *Flu* is not included and projection takes place on the disease) is generalized to be $\pi_{D1}(\rho_{N1\leftarrow Name, D1\leftarrow Disease}(Ill))$ by dropping the second occurrence of *Ill* in the join and consequently removing the selection condition for *Cough*, we could assign similarity degrees by comparing each disease with *Cough* as follows. Assuming that $sim(Cough,Cough)=1$, $sim(Cough,Flu)=0.8$, $sim(Cough,BrokenLeg)=0.4$, and $sim(Cough,Sinusitis)=0.7$, we obtain the following answer table:

| D1 | |
|---|---|
| Cough | 1.0 |
| BrokenLeg | 0.4 |
| Sinusitis | 0.7 |
| Flu | 0.8 |

If more than one selection condition is dropped, that is, if more than one constant $a_j$ is affected, the degrees for all these constants could be aggregated into a unique degree for the answer tuple.

## 3.3 Goal replacement

Goal Replacement can be treated with methods similar to the Dropping Condition case as with GR also some of the relations are removed from the join expression and hence also selection or equality conditions or projection attributes may be affected. However, we could consider as more positive the case that attributes are replaced with GR (instead of totally dropped with DC). For example, as the selection condition for *Flu* is replaced by one for *Inhaler* with the example rule, this would not be penalized as in the DC case.

A semantic enhancement of GR could obtain a similarity degree by aggregating similarity values for the constants in the replacement rule. For the example rule, assuming that $sim(Flu,Inhaler)=0.5$, the answer tuples obtained after GR would all have this similarity degree.

## 4. CONCLUSION

This article presented a collection of heuristics to determine relevancy of answer tuples via a similarity degree after a failing query has been modified (more precisely, generalized) to obtain informative answers. Such a notion of relevancy can be used to improve cooperativeness of database systems in the sense that they support the user in finding answers that do not match his original query exactly but might nevertheless be relevant for him. Applicability of the heuristics may highly depend on the context in which a database is used and good enough similarity values for the particular application context are crucial.

Future work will address a formalization of the approach with the help of fuzzy logic by extending the work of Behlolavek et al (2011). In particular, we aim at being able to obtain similarity degrees for answer tuples obtained by a combination and iteration of generalization operators.

## ACKNOWLEDGEMENT

I thank Maheen Bakhtyar, Katsumi Inoue, Robert Kowalski and Lucie Urbanova for helpful discussions and inspiring ideas.